\documentclass{aa}  
\usepackage{graphicx}
\usepackage{txfonts}
\usepackage{bm}
\usepackage{amsmath}
\usepackage{siunitx}
\usepackage{xcolor}
\usepackage{hyperref}

\bibpunct{(}{)}{;}{a}{}{,} 
\usepackage{bbold}
\hypersetup{colorlinks=true,citecolor=blue,linkcolor=blue,urlcolor=blue}
\usepackage{tabularx}
\usepackage{xspace}

\usepackage{soul}
\usepackage{fontawesome}
\usepackage{ragged2e}

\newcommand*{\Euclid}{\textit{Euclid}\xspace}
\newcommand*{\LSST}{\textit{LSST}\xspace}

\definecolor{darkraspberry}{rgb}{0.53, 0.15, 0.34}

\newcommand{\vilasini}[1]{#1}

\definecolor{color1}{rgb}{0.3, 0.7, 0}

\setstcolor{magenta}
\newcommand{\strike}[1]{}
\begin{document} 
   \title{Theoretical wavelet $\ell_1$-norm from one-point PDF prediction}

   \author{Vilasini Tinnaneri Sreekanth
          \inst{1}
          \and
          Sandrine Codis\inst{1}
          \and
          Alexandre Barthelemy\inst{2}
          \and
          Jean-Luc Starck
          \inst{1,3}
          }

   \institute{Université Paris-Saclay, Université Paris Cité, CEA, CNRS, AIM, 91191, Gif-sur-Yvette, France\\
              \email{vilasini.tinnanerisreekanth@cea.fr}
        \and
        Universit\"{a}ts-Sternwarte, Fakult\"{a}t f\"{u}r Physik, Ludwig-Maximilians-Universit\"{a}t M\"{u}nchen, Scheinerstra{\ss}e 1, 81679 M\"{u}nchen, Germany 
         \and
             Institutes of Computer Science and Astrophysics, Foundation for Research and Technology Hellas (FORTH), Greece       
             }

   \date{Received ?; accepted ?}
 
  \abstract
   {Weak gravitational lensing, resulting from the bending of light due to the presence of matter along the line of sight, is a potent tool for exploring large-scale structures, particularly in quantifying non-Gaussianities. It stands as a pivotal objective for upcoming surveys. In the realm of current and forthcoming full-sky weak-lensing surveys, the convergence maps, representing a line-of-sight integration of the matter density field up to the source redshift, facilitate field-level inference, providing an advantageous avenue for cosmological exploration. Traditional two-point statistics fall short of capturing non-Gaussianities, necessitating the use of higher-order statistics to extract this crucial information. Among the various higher-order statistics available, the wavelet $\ell_1$-norm has proven its efficiency in inferring cosmology (\cite{2021A&A...645L..11A}). However, the lack of a robust theoretical framework mandates reliance on simulations, demanding substantial resources and time. }
   {Our novel approach introduces a theoretical prediction of the wavelet $\ell_1$-norm for weak lensing convergence maps, grounded in the principles of Large-Deviation theory. This method builds upon recent work by \cite{2021MNRAS.503.5204B}, offering a theoretical prescription for aperture mass one-point probability density function. }
   {We present, for the first time, a theoretical prediction of the wavelet $\ell_1$-norm for convergence maps, derived from the theoretical prediction of their one-point probability distribution. Additionally, we explore the cosmological dependence of this prediction and validate the results on simulations.\href{https://github.com/vilasinits/LDT_2cell_l1_norm}{\faGithub}}
   {A comparison of our predicted wavelet $\ell_1$-norm with simulations demonstrates a high level of accuracy in the weakly non-linear regime. Moreover, we show its ability to capture cosmological dependence, paving the way for a more robust and efficient parameter inference process.}
   {}


   \maketitle

%
%
\section{Introduction}
\label{sec: Introduction}
Our present comprehension of the genesis of large-scale structure \citep{1980lssu.book.....P} posits that it emerged through gravitational instability driven by primordial fluctuations in matter density. Realistic models detailing structure formation prescribe an initial spectrum of perturbations reflecting the primordial spectrum, characterised by small fluctuations at large scales. In an LCDM Universe, the variance of the density fluctuations, denoted by $\sigma(R)^2$, is inversely proportional to (some power of) scales so that at small scales, the variance is large and non-linear effects become important.

Hence, there are two limiting regimes: the linear regime which is characterized by $\sigma^2(R) << 1$, and the non-linear regime which is given by $\sigma^2(R) >>1$.  In particular, if the primordial density fluctuations are Gaussian, then they remain Gaussian in the subsequent linear regime, and Fourier modes evolve independently. However, the coupling between different Fourier modes becomes significant and plays a pivotal role in modifying the statistical properties, manifesting as higher-order connected correlations in the non-linear regime \citep{Bernardeau_2002}. 

One of the powerful methods, to probe these non-linearities present in the large-scale structure (hereafter LSS) is to look at the distortions of the distant images of galaxies. Distortions happen because of gravitational lensing, which is the phenomenon in which the paths of the photons are bent due to the presence of massive objects such as galaxies or galaxy clusters. 
Weak gravitational lensing denotes that, except for rare cases of strong lensing, these distortions are often subtle, requiring statistical analysis over a vast number of galaxies to detect a signal. Analyzing these distortions provides a unique opportunity to investigate the distribution of matter in the Universe and infer cosmological parameters. For a more comprehensive review of weak lensing, readers are referred to (\citet{Mandelbaum_2018};\citet{kilbinger2018cosmological}), and \citet{Kilbinger_2015}.

In this era of precision cosmology, with past, present, and future surveys \strike{like} \vilasini{such as the} Canada-France-Hawaii Telescope Lensing Survey (CFHTLenS\vilasini{)} \citep{2012MNRAS.427..146H}, Hyper Suprime\vilasini{-}Cam (HSC) \citep{mandelbaum2017weak}, \Euclid \citep{laureijs2011euclid}, \LSST \vilasini(\citep{2019ApJ...873..111I}), to name a few, we now have access to the non-Gaussian part of cosmological signals, which are induced by the non-linear evolution of the structures on small scales. All of these experiments, consider weak gravitational lensing to be a key probe -- jointly with galaxy clustering -- to explore the current unanswered questions in Cosmology like the neutrino mass sum \citep{Lesgourgues_2012, 2019PhRvD..99f3527L}, the nature of dark energy and dark matter \citep{Huterer_2010}, and it offers substantial constraints on standard cosmological parameters like the mean matter density, the amplitude of matter fluctuations \citep{TROXEL20151}. 

The conventional approach to infer the cosmology from the data involves the computation of the two-point statistics, which has been employed with remarkable success in the past \citep{2008PhR...462...67M, Kilbinger_2015, 2017SchpJ..1232440B,2017MNRAS.465.1454H} \vilasini{(\citep{2024A&A...684A.138E, Loureiro_2023, 10.1093/mnras/stac046})}. However, it is not sufficient if we want to probe the non-Gaussianities present \citep{weinberg2013observational}. This limitation arises from the construction of the power spectrum, which considers information solely from the norm of wave-vectors while neglecting phase information, thereby discarding a significant aspect of structural details. Consequently, there arises a necessity to complement this approach with an alternative higher-order statistic capable of effectively capturing the non-Gaussian features embedded in the structure. Higher-order statistics such as peak counts \citep{1999MNRAS.302..821K,liu2015cosmology,liu2015cosmological,2015A&A...583A..70L,2017A&A...599A..79P,2020PhRvD.102j3531A}, higher moments \citep{petri2016mocking,2018A&A...619A..38P,2020MNRAS.498.4060G}, Minkowski functional \citep{2012PhRvD..85j3513K,2020A&A...633A..71P}, three-point statistics \citep{takada2004cosmological,semboloni2011weak,2019MNRAS.490.4688R} and wavelet and scattering transform \citep{2021A&A...645L..11A,Cheng2021} amongst many others, can better probe the non-Gaussian structure of the Universe and provide additional constraints on the \vilasini{cosmological} parameters. Another example was given in \citet{2021A&A...645L..11A} employing the \strike{so-called} starlet $\ell_1$-norm that has the advantage of being easy to measure and was claimed to encompass even more cosmological information than the power spectrum or peak and void counts in the setting considered. The $\ell_1$-norm of a starlet, which is a type of wavelet that uses $B3$-spline, offers an efficient multiscale computation of all map pixels including the under-densities and over-densities distribution, partially probing similar information to peak and void counts.

The Probability Distribution Function (hereafter PDF) of the weak lensing convergence map ($\kappa$) serves as another valuable repository of cosmological information that has drawn a lot of interest in recent years, with many theoretical and numerical works including \citep{bernardeau2001construction,liu2019constraining,2021MNRAS.503.5204B}. The convergence field $\kappa$ represents the weighted projection of matter density fluctuations along the line of sight and its higher-order correlations offer a promising avenue for addressing challenges inherent in standard weak lensing analyses, particularly those related to degeneracies in the two-point correlation function (2PCF).

The one-point $\kappa$-PDF statistic, obtained by quantifying smoothed $\kappa$ field values within predefined apertures or cells, presents a practical advantage as its measure is rather simple. This simplicity contrasts with other non-Gaussian probes used for studying the weak lensing convergence field, such as the bispectrum (which involves counting triangular configurations) or Minkowski functionals (a topological measurement).
Previous research has successfully devised a precise theoretical model, grounded in large deviation theory (LDT), for both the cumulant generating function and the \strike{probability distribution function} \vilasini{PDF} of the lensing $\kappa$ field as well as the aperture mass \citep{2021MNRAS.503.5204B}. Notably, this one-point PDF could be directly linked to the wavelet coefficients at different scales \citep{2021A&A...645L..11A}.

Let us emphasize that similarly to the convergence PDF, all of the above-mentioned non-Gaussian statistics, which probe deviations from Gaussian behaviour, are also commonly computed using the convergence maps. However, these convergence maps are not observed directly and are reconstructed from the reduced shear maps. The shear maps since its first detection two decades ago \citep{2000MNRAS.318..625B,kaiser2000largescale,vanwaerbeke2000detection} have been a powerful cosmological probe. However, due to its spin-2 nature, it is rather difficult to obtain higher-order summary statistics from shear. While the convergence maps in principle contain the same information as the shear maps \citep{2002A&A...396....1S,2011A&A...533A..48S}, the compression of the lensing signal is greater in convergence maps compared to the shear field, resulting in easier extraction and reduced computational costs, with the caveat that the reconstruction of the convergence maps is not perfectly solved and is a very complex ill-defined inverse problem. Convergence maps emerge as a novel tool, potentially offering additional constraints that complement those derived from the shear field. However, accessing this information is not straightforward and implies the use of a reconstruction method (or mass inversion). In particular, let us note that due to the non-Gaussian nature of the weak-lensing signal at small scales, employing mass-inversion methods with smoothing or de-noising for regularization may not be optimal \citep{starck21:mcalens,jeffrey20:deepmass}. 

The objective of this paper is to present, for the first time, the prediction of the wavelet $\ell_1$-norm derived from theoretical predictions of the \strike{Probability Distribution Function} \vilasini{PDF} of convergence maps. The paper is organized as follows: In Sect. \ref{sec convergence maps}, we begin by revisiting weak lensing convergence, followed by an introduction to the wavelets and wavelet $\ell_1$-norm in Sect. \ref{sec: introduction to wavelet and wavelet l1}. This is followed by the introduction of the LDT formalism for the $\kappa$-PDF and extending this to the wavelet coefficients, in Sect. \ref{sec: LDT and extension to wavelet coefficients}. Section \ref{sec: deriving l1 norm from pdf} delves into the derivation of this wavelet $\ell_1$-norm for the wavelet coefficients from theory. Subsequently, we present and discuss the results in Sect. \ref{sec: results}, concluding with the summarizing of our findings in Sect. \ref{sec: discussion  and conclusion}. 

\section{Wavelet $\ell_1$-norm: definition and measurements}

We introduce in this section first the expression for the convergence maps and move to a more general definition of the wavelets and the wavelet $\ell_1$-norm and see how it is related to the PDF.
\label{sec: introduction to wavelet and wavelet l1}

\subsection{Convergence maps}
\label{sec convergence maps}

Let us start with the expression for convergence, which is given by a projection of density along the comoving coordinates, weighted by a lensing kernel involving the comoving distances. It is given by \citep{1999ARA&A..37..127M}
\begin{equation}
    \kappa(\bm{\vartheta}) = \int_0^{\chi_s} {\rm d}\chi \omega(\chi,\chi_s) \delta(\chi,\mathcal{D}\bm{\vartheta}),
    \label{def-convergence}    
\end{equation}
where $\chi$ is the comoving radial distance -- $\chi_s$ the radial distance of the source -- that depends on the cosmological model, and $\mathcal{D}$ is the comoving angular distance given by
\begin{equation}
    \mathcal{D}(\chi) \equiv\left\{
    \begin{aligned}{\frac{\sin (\sqrt{K} \chi)}{\sqrt{K}}}  {\text { for } K>0} \\ {\chi \qquad}  {\text { for } K=0} \\ {\frac{\sinh (\sqrt{-K} \chi)}{\sqrt{-K}}}  {\text { for } K<0}
    \end{aligned}\right. ,
\end{equation}
with $K$ the constant space curvature. The lensing kernel $\omega$ is defined as
\begin{equation}
    \omega(\chi,\chi_s) = \frac{3\,\Omega_m\,H_0^2}{2\,c^2} \, \frac{\mathcal{D}(\chi)\,\mathcal{D}(\chi_s-\chi)}{\mathcal{D}(\chi_s)}\,(1+z(\chi)),
    \label{eq: lensing kernel}
\end{equation}

\vilasini{where, $c$ is the speed of light, $\Omega_m$ the matter density parameter and $H_0$ is the values of bubble constant at redshift $z=0$.}
The shear field $\gamma(\theta)$ which is interpretable directly from the observation is related to the weak lensing convergence $\mathcal{\kappa}$ through mass inversion \citep{starck2006,2015A&A...581A.101M}. The convergence mass maps are in principle constructed only up to a mass sheet degeneracy. However, constructing a mass map by convolving the $\kappa$ map with a radically symmetrical compensated filter, makes the statistics insensitive to mass sheet degeneracy. 

\subsection{Wavelets}
\label{sec: wavelet}
A wavelet transform enables us to decompose an image $\kappa$ into different maps called wavelet \strike{scales} \vilasini{coefficients} $w_{\theta_{j}}$, i.e.  ${\cal W} \kappa = \left\{w_{\theta_1}, \dots, w_{\theta_j}, \dots w_{\theta_J } \right\}$ where $\theta_j$ is the angular scale, and $\theta_{J}$ is the largest scale used in the analysis.The wavelet coefficients are the values of the wavelet scales at the pixel coordinates (x, y).
Considering a wavelet function $\Upsilon$, and noting $\Upsilon_{\theta_j,x,y}(m,n) =  \Upsilon \left( \frac{  m - x}{\theta_j}, \frac{ n  - y}{\theta_j} \right)$ the dilated wavelet function at scale $\theta_j$ and at spatial location $(x,y)$, the wavelet coefficients $w_{\theta_j}$ are computed as the inner products 
\begin{eqnarray}
    w_{\theta_j}(x,y) & = & <\kappa, \Upsilon_{\theta_j,x,y}> \\ \nonumber     
                      & = &   \sum_m \sum_n \kappa(m,n)  \Upsilon \left( \frac{  m  - x}{\theta_j}, \frac{  n  - y}{\theta_j} \right), 
    \label{eq: wavelet convolution}
\end{eqnarray}
{The wavelet function $\Upsilon$ is often chosen to be derived from the difference of two resolutions, for example as $\Upsilon(x,y) = 4 \xi(2x,2y) - \xi(x,y) $ as utilized in the starlet wavelet transform \citep{book}, where the function $\xi$ is called the scaling function, typically a low-pass filter. In our case, we write the wavelet coefficients as
\begin{equation}
    w_{\theta_j}(x,y) = <\kappa,\xi_{\theta_{j+1},x,y}> - <\kappa,\xi_{\theta_{j},x,y}>,
    \label{eq: wavelet coefficients}
\end{equation}
with $\xi_{\theta_j,x,y}(m,n) =  \xi \left( \frac{  m - x}{\theta_j}, \frac{ n  - y}{\theta_j} \right)$.
Employing dyadic scales, i.e. $\theta_j = 2^{j-1} \theta_1$, enables us to have very fast algorithms through the use of a filter bank. See \citet{book} for more details. 

A wavelet acts as a mathematical function localized in both the spatial and the Fourier domains, thus suitable for analysing the lensing signal's structures at various scales. 
An important advantage of wavelet analysis, compared to a standard multi-resolution analysis through the use of a set of Gaussian functions of different sizes, is that it decorrelates the information. As an example, \citet{2023A&A...672L..10A} have shown that the covariance matrix derived from a wavelet peak count analysis was almost diagonal, and neglecting the off-diagonal terms has little impact on the cosmological parameters estimation. This would not be the case with a multi-scale Gaussian analysis. Other advantages are that very fast algorithms exist allowing us to compute all scales with a low complexity, and also that wavelet functions are exactly equivalent to traditional weak-lensing aperture mass functions which have been used for decades \citep{Leonard_2012} (aperture mass is just the convergence/shear within a compensated filters, which is one property of wavelets). Several statistics have been derived from wavelet coefficients in the past, such as cumulants (up to order 6) \citep{7026230} or peak counts \citep{2020PhRvD.102j3531A}. It has recently been shown \citep{2021A&A...645L..11A} that the $\ell_1$-norm of the wavelet scales is very efficient in constraining cosmological parameters. In \cite{2023A&A...672L..10A}, a toy model was utilized, incorporating a mock source catalogue for weak lensing and a mock lens catalogue for galaxy clustering sourced from the cosmo-SLICS \vilasini{(\citep{2019A&A...631A.160H})} simulations to mimic the KiDS-1000 data survey properties described in \citet{10.1093/mnras/sty2271} and \citet{2021A&A...647A.124H}. The study examined forecasts concerning the matter density parameter $\Omega_m$, the matter fluctuation amplitude $\sigma_8$, the dark energy equation of state $w_0$, and the reduced Hubble constant $h$. It was observed that the wavelet $\ell_1$-norm demonstrated superior performance compared to peaks, multi-scale peaks, or a combination thereof \citep{2023A&A...672L..10A}. Therefore, we propose to build the theory for wavelet $\ell_1$-norm, which could be used for constraining the cosmological parameters with upcoming surveys.

\subsection{Wavelet $\ell_1$-norm}

To measure the wavelet $\ell_1$-norm from a $\kappa$ map, we first obtain the wavelet scale by convolving the map with the wavelet function, as given in Eq. \ref{eq: wavelet convolution}. 

We then compute a histogram of the values of the wavelet scale at each pixel coordinate, $w_{\theta_j}(x,y)$,  using a specific binning \vilasini{with bin edges denoted} $\{B_i\}_{1\leq i\leq N}$. 

One can then obtain the {\it normalised}\footnote{In the literature, the common practice has been to use an un-normalised version of the $\ell_1$-norm, but for an easier comparison with prediction and simulation we choose here to normalize this quantity. For the sake of simplicity, we will call it $\ell_1$-norm in this paper. } wavelet $\ell_1$-norm by extracting the $\ell_1$-norm of the histogram such that,

\begin{equation}
    \ell_1^{\theta_j,i} = \sum_{k=1}^{\# coef(S_{\theta_j,i})} |S_{\theta_j,i}[k]| ~/ N / \Delta B,
    \label{eq: l1 norm}
\end{equation}
where the set of coefficients at scale ${\theta_j}$ and amplitude bin $i$, 
$S_{\theta_j,i} = \left\{w_{\theta_j}~  / ~~  w_{\theta_j}(x,y)  \in [ B_i,  B_{i+1} ] \right\}$, 
depicts the wavelet coefficients $w_{\theta_j}$ having an amplitude within the bin $ [B_i,  B_{i+1}]$, and the pixel indices are given by $(x,y)$ and $N$ is the total number of pixels and $\Delta B$ is the uniform bin width.
In other words, for each bin, the number of pixels $k$, that fall in the bin is collected and the absolute values of these pixels are summed to obtain the $\ell_1$-norm at this bin $i$. 

As we already highlighted before, this definition enables us to gather the data represented by the absolute values of every pixel of the map, rather than solely describing it through the identification of local minima or maxima. It has the advantage of being a multiscale approach. In addition, the robustness of $\ell_1$-norm statistics is also well established for decades in the statistical literature \citep{huber1987place,gine2003bm}.

It turns out that the $\ell_1$-norm of the wavelet coefficients can be related directly to the PDF of the wavelet coefficients via the following relation
\begin{equation}
    \ell_1^{\vilasini{\theta_j,}i} = P(w_j)^i \times |B^i|,
    \label{eq: l1 pdf relation}
\end{equation}

which states that the $\ell_1$-norm \vilasini{for a given scale j and} at a given bin $i$ ($B_i$) can be related to its PDF by multiplying the PDF value of \strike{at} the bin $i$ by the absolute value of the bin $i$.

\subsection{Gaussian $\ell_1$-norm} 
Let us now look at the anticipated wavelet $\ell_1$-norm expected from a Gaussian distribution. The generic shape of a wavelet $\ell_1$-norm obtained from a Gaussian PDF 
\begin{equation}
    P(x) = \frac{1}{\sigma \sqrt{2\pi}} \exp\left(-\frac{1}{2}\left(\frac{x-\mu}{\sigma}\right)^2\right)
    \label{eq:gaussian}
\end{equation}
with mean $\mu = 0$ and $\sigma = 0.1$ using the relation shown in Eq. \eqref{eq: l1 pdf relation} is demonstrated in Fig. \ref{fig: l1 for gaussian}.

\begin{figure*}
   \resizebox{\hsize}{!}
    {\includegraphics{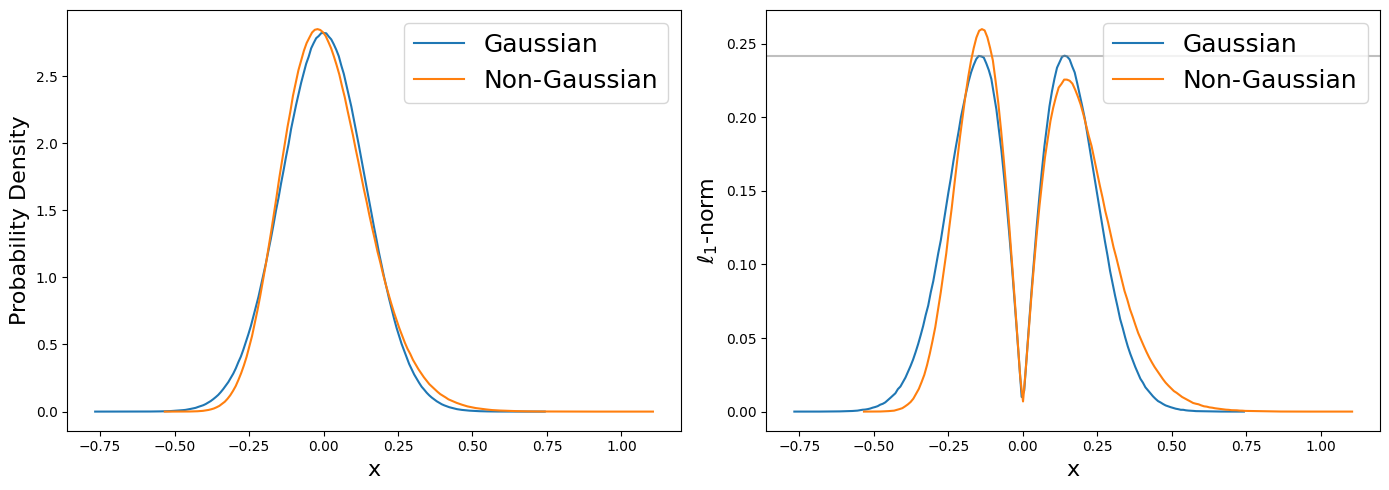}}
    \caption{This figure illustrates the PDF and the $\ell_1$-norm for a Gaussian distribution in blue and the non-Gaussian distribution in orange. On the left, we present the Probability Distribution Function (PDF) for the two distributions, and on the right, we display the derived $\ell_1$-norm of these PDFs. Let us note that the heights of the peaks are the same for a Gaussian distribution but this does not hold for a non-Gaussian PDF.}
    \label{fig: l1 for gaussian}
    
\end{figure*}

Here, we see that the height of the two peaks obtained for the wavelet $\ell_1$-norm for the Gaussian (blue solid line) distribution are equal as expected. 
But, as the Universe evolves, non-linearities develop, which means that the convergence field $\kappa$ that we are interested in can no longer be modelled using a Gaussian model. These non-linearities would lead to asymmetry which would also be evident from the peaks of the wavelet $\ell_1$-norm, which will no longer be corresponding. This is demonstrated in Fig. \ref{fig: l1 for gaussian} for a test case scenario, where we have applied a non-linear exponential transformation, to the Gaussian field to obtain a non-Gaussian field (orange solid line) and obtained the PDF and the $\ell_1$-norm. 
Here we would like to emphasize that the usual way to extract this non-linear information is either by relying on heavy $\mathcal{N}$-body simulations or by using an approximation, as is done in \citep{Tessore_2023}. But, thanks to \strike{Large Deviation Theory} \vilasini{LDT}, we now have a theoretically motivated way to obtain the PDF in the mildly non-linear regime, using which we will show in this paper that we can also obtain the wavelet $\ell_1$-norm in the mildly non-linear regime. 

\section{Large Deviation Theory formalism}
\label{sec: LDT and extension to wavelet coefficients}

Let us now, look at the application of the \strike{Large Deviation Theory Formalism (LDT)} \vilasini{LDT} to the convergence field. LDT, as explored in earlier works \citep{doi:10.1137/1.9781611970241}, examines the rate at which the probabilities of specific events exponentially decrease as a key parameter of the problem undergoes variation. To see how to derive the wavelet $\ell_1$-norm based on the LDT formalism, we first need to recap how to obtain the PDF from the LDT formalism.

\subsection{LDT for the matter field}
The application of LDT to the field of Large Scale Structure cosmology has been systematically developed in recent years and will be applied in the specific context of cosmic shear observations in this work. \citep{2016PhRvD..94f3520B} clarified the application of the theory to the cosmological density field, establishing its link to earlier studies focused on cumulant calculations and modelling the matter \strike{Probability Distribution Function (} PDF \strike{)} through perturbation theory and spherical collapse dynamics \citep{2002A&A...382..412V,PhysRevD.90.103519}. \cite{2020MNRAS.492.3420B} then provided an LDT-based prediction for the top-hat-filtered weak-lensing convergence PDF on mildly nonlinear scales, building upon earlier work by \citep{2000A&A...364....1B}.

A more detailed derivation of the specific equations used is presented in the Appendix \ref{appendix:a} for the interested readers. In this section, we simply recall the main equation that will then be needed for us to derive the wavelet $\ell_1$-norm prediction.

Within the context of the application of LDT in Cosmology, there are three main steps involved. The first is the derivation of the rate function, which is obtained directly from the first principles of Cosmology. Using the contraction principle in the LDT formalism, we can connect the statistics of the late-time non-linear densities to the earlier time density field as soon as the most likely mapping between the two is known. Previous works have shown that assuming spherical collapse for this mapping provides us with a very accurate prediction for cosmic field PDF (density, velocity, convergence, ...). 

In particular, the result for convergence reads
\begin{equation}
    P(\kappa) = \int_{-i\infty}^{+i\infty} \frac{d\lambda}{2\pi i} \exp(-\lambda \kappa + \phi_{\kappa,\theta}(\lambda))
    \label{eq: inverse laplace}
\end{equation}
where, $P(\kappa)$ is the PDF of the convergence field $\kappa$, and $\phi_{\kappa,\theta}$ is the cumulant generating function (CGF) of the field $\kappa$ at angular scale $\theta$. The derivation of the CGF in the above equation and PDF from there is explained in more detail in Appendix \ref{appendix:a}.

\subsection{Extending to wavelet coefficients}

From the PDF, we can now compute the $\ell_1$-norm.
As was given in the Sect. \ref{sec: introduction to wavelet and wavelet l1}, a wavelet can be written as a function of scaling functions. Though various wavelet filters exists in literature, in this study, we employ a function of concentric disks to construct the wavelet filter $\Upsilon_{\theta_1}$.

The scaling function $\xi$ is then given by a spherical top hat filter and is given by 
\begin{equation}
    \hat{\xi}_{\theta_i} = 2 J_1(\theta_il)/(\theta_i l)
    \label{eq: TH filter}
\end{equation}
 where $J_1$ is the first Bessel function of the first kind. This is applied, as shown in Eq. \eqref{eq: wavelet coefficients}, to obtain our compensated filter.
In Fig. \ref{fig: filter}, we show the compensated filter (black solid line) that is used in this work, which is given as a function of two scaling functions (orange and blue solid lines) at different scales. 

\begin{figure}
   \resizebox{\hsize}{!}
            {\includegraphics{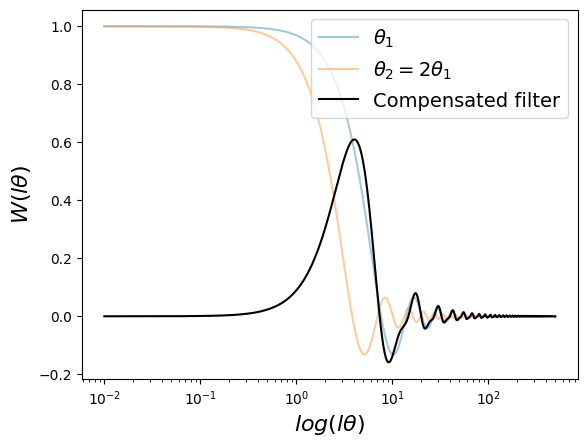}}
    \caption{The compensated filter (depicted by the black solid line) is derived through a difference of two top-hat filters at different scales, as described in Eq. \eqref{eq: wavelet coefficients} and Eq. \eqref{eq: TH filter}. The blue and orange solid lines represent the individual spherical filters obtained at radii $\theta_1$ and $\theta_2 = 2\theta_1$, respectively. For visualization purposes, the compensated filter is multiplied by $-1$.}
    \label{fig: filter}
\end{figure}

We now apply the LDT formalism described previously \strike{for} \vilasini{to} the wavelet coefficients described in Eq. \eqref{eq: wavelet coefficients} and obtain the PDF of the wavelet coefficients ($w_{\theta_1}$)  by using the scales $\theta_1, \theta_2=2 \theta_1$ as 
\begin{equation}
    P(w_{\theta_1}) = \int_{-i\infty}^{+i\infty} \frac{d\lambda}{2\pi i} \exp \left(-\lambda w_{\theta_1} +  \phi_{w_{\theta_1}}(\lambda)\right).
    \label{eq: PDF wavelet coefficients}
\end{equation}

We refer the readers to Appendix \ref{appendix:a}, for more details on the derivation of the CGF in the above equation and the resulting PDF. We emphasize that $w_{\theta_1}$ is \strike{literally just the regular} \vilasini{essentially the standard aperture mass} aperture mass defined on the convergence field using a difference of two top-hat filters, as was discussed previously.

\subsection{Wavelet $\ell_1$-norm predicted via \vilasini{L}arge \vilasini{D}eviation \vilasini{T}heory}
\label{sec: deriving l1 norm from pdf}

In the previous sections, we saw how the LDT formalism can be applied to derive the PDF of the wavelet coefficients. We will now see how to derive the wavelet $\ell_1$-norm from the PDF of the wavelet coefficients $P(w_{\theta_1})$. This wavelet $\ell_1$-norm given in Eq. \eqref{eq: l1 norm}, which is the sum of the absolute values of the pixels in each bin, is \strike{akin} \vilasini{equivalent} to multiplying the counts/PDF value by the absolute value of the bin it corresponds to, as is given in Equation \eqref{eq: l1 pdf relation}. Extending this to the PDF of the wavelet coefficients from the LDT formalism, we get the final equation for bin $i$
\begin{equation}
   \ell_1^{i}  = P^i (w_{\theta_1}) \times |w_{\theta_1}^i|.
   \label{eq: derived l1 norm}
\end{equation}

As a first approximation, one can assume that the PDF is Gaussian, which is the case when we consider large scale or early time in the standard model of Cosmology. In the subsequent section, we look at obtaining the $\ell_1$-norm for a Gaussian case first and then highlight the motivation to go beyond Gaussian models.

\section{Confronting $\ell_1$-norm prediction to simulations}
\label{sec: results}

\subsection{Measurements from simulation}
\label{sec: takahashi simulation}

To demonstrate the accuracy of the theory for realistic surveys, we compare our prediction results with full-sky simulations of \citet{Takahashi_2017} to avoid additional errors from small-scale patch sizes. The simulations of \citet{Takahashi_2017} provides full sky lensing maps at a fixed cosmology \footnote{\url{http://cosmo.phys.hirosaki-u.ac.jp/takahasi/allsky_raytracing/}} each for two grid resolutions: $4096 ^2$ and $8192^2$. The data sets include the full sky maps at intervals of $150 h^{-1}$ comoving radial distance from redshifts $z = 0.05$ to $5.3$. We use the simulated convergence maps that are directly provided as products of these simulations. The smoothing of the convergence maps is implemented as a direct convolution of the convergence map with the wavelet filter at a given angular scale. Once the filtered field is obtained, we measure the sum of the absolute values of the pixels of the smoothed $\kappa$ mass map in linearly spaced bins of the range of $\kappa$ to obtain the wavelet $\ell_1$-norm.

\subsection{Obtaining the prediction}

To obtain the prediction of the wavelet $\ell_1$-norm for each of the cosmologies used here, we first use CAMB  \citep{2011ascl.soft02026L} to get the linear and non-linear(Halofit \vilasini{takahashi \citep{2012ApJ...761..152T}}) matter power spectra. We also use CAMB to obtain the comoving radial distances $\chi(z)$. The density CGF (Eq. \eqref{eq: CGF}) is calculated for redshift slices between $z=0$ and the source redshift $z_s$ using the full Halofit power spectrum as input. These projected CGFs are then rescaled by the measured variance $\sigma^2_{M_{ap},sim}$ (see Appendix \ref{appendix:a} for more details), before going through an inverse Laplace transform to get the final PDF. This predicted PDF is used to derive the predicted wavelet $\ell_1$-norm as given in Eq. \eqref{eq: derived l1 norm}.

\subsection{Validating the prediction with simulation}

In Figures \ref{fig: different z} and \ref{fig: different r}, we show the predicted wavelet $\ell_1$-norm and the one measured from the simulated convergence map for different source redshifts $z_s \approx 1.2, 1.4, 2.0$ at $\theta =  \SI{20}{\arcminute}$ and for different scales $\theta = \SI{15}{\arcminute}, \SI{18}{\arcminute}, \SI{20}{\arcminute}$ at source redshift $z_s = 1.423$ respectively. On the bottom panels, we show the residuals. For comparison, we display the PDF and the residuals of a Gaussian distribution (dash-dot line) with the same mean and variance as the simulation. The Gaussian PDF is obtained using Eq. \eqref{eq:gaussian}, and the $\ell_1$-norm is derived as explained previously in Eq. \eqref{eq: derived l1 norm}. In both plots, the $x$-axis for the residuals is scaled by the standard deviation. The shaded regions in the colors correspond to the $3\sigma$ region obtained from the error bars. 

The error bars are obtained by taking the standard deviation of the values of the wavelet $\ell_1$-norm for 10 different patches of the full sky Takahashi map in each of the bins.

Let us first concentrate on the Gaussian curves. We see a clear mismatch with the Gaussian prediction as expected since the measurements show a clear asymmetry unless the Gaussian model which by definition is symmetric.  

Let us now focus on the LDT prediction that goes beyond the Gaussian regime. In this case, we observe a better agreement between the simulation and the predicted wavelet $\ell_1$-norm results to percent levels (which corresponds to the simulation error bars in shaded area) within the $2-\sigma$ region around zero on the $x$-axis. This demonstrates the predictive accuracy of the results derived from the LDT formalism. It is important to note that this $2-\sigma$ specification refers to the range on the $x$-axis, rather than representing error bars around the mean signal value.

In these plots, we indeed clearly see that LDT is capable of capturing the asymmetry of the two peaks hence valuable information about the non-gaussianities. To better assess this aspect, we also show, in Table \ref{tab:moments}, the cumulants obtained for the three scales at source redshift $z_s = 1.423$. The cumulants shown here are the variance ($\sigma^2$) and reduced skewness ($S_3$) which, given that the mean convergence is zero, are obtained as 
\begin{align}
    \sigma^2 &= E\left[\kappa^2\right]\\
    S_3 &= \frac{E[\kappa^3]}{\sigma^4}
\end{align}

where,
\begin{equation}
    E[\kappa^n] = \int P(\kappa)\kappa^n d\kappa.
\end{equation}

 \begin{table}
      \caption[]{The cumulants (standard deviation $\sigma$, skewness $S_3$) obtained from the PDF from the LDT prediction and the values obtained from the Takahashi simulation at source redshift $z_s \approx 2.05$ for scales 15, 18, and 20 arcminutes. The corresponding PDFs are shown in Figure \ref{fig: different r}}
         \label{tab:moments}
        $$
         \begin{array}{|p{0.25\linewidth}|l|c|c|c|r|}
            \hline
            Data & \sigma^2 \times 10^{-5} & S_3 \\
            \hline
            Prediction 15' & 2.92 &-26.3  \\
            \hline
            Simulation 15' & 2.92 & -42.8 \\
            \hline
            Prediction 18' & 2.61 & -25.94 \\
            \hline
            Simulation 18' & 2.61 & -38.2 \\
            \hline
            Prediction 20' & 2.44 & -25.7 \\
            \hline
            Simulation 20' & 2.44 & -35.4 \\
            \hline
         \end{array}
        $$
   \end{table}

    \begin{figure}
    \resizebox{\hsize}{!}
        {\includegraphics{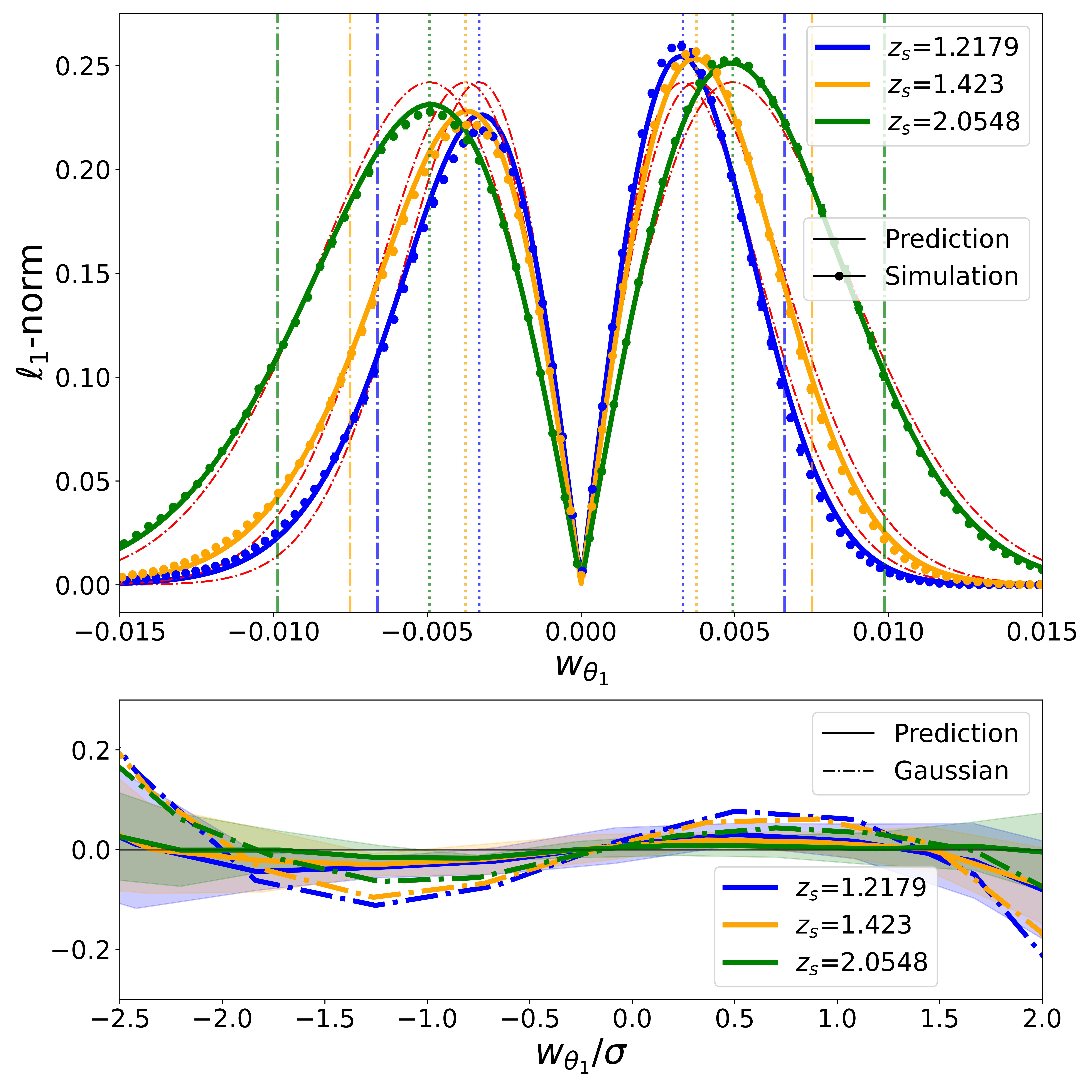}}
      \caption{Top panel: Predicted (solid) $\ell_1$-norm as compared to measurements in simulation (dot markers) for an inner radius $\theta_1 = \SI{20}{\arcminute}$ and different source redshifts $z_s = 1.21,1.43,2.05$, displayed with blue, orange, and green lines respectively. Red dash-dotted lines show the Gaussian prediction for reference. \vilasini{The vertical dotted and dot-dash lines correspond to the $1\sigma$ and $2\sigma$ regions around the mean of the $w_{\theta_1}$ for each of the case considered.}
      Bottom panel: residual of the prediction relative to the simulation (dotted lines). For reference, the dash-dotted plots illustrate the residual of the $\ell_1$-norm derived from the Gaussian PDF with the same mean and variance as the simulation PDF. The shaded region indicates the $3\sigma$ region around the error bars for each redshift. The prediction is in good agreement with the measurements up to approximately 2 sigma and remains within percent levels.}
        \label{fig: different z}
   \end{figure}

   \begin{figure}
   \resizebox{\hsize}{!}
            {\includegraphics{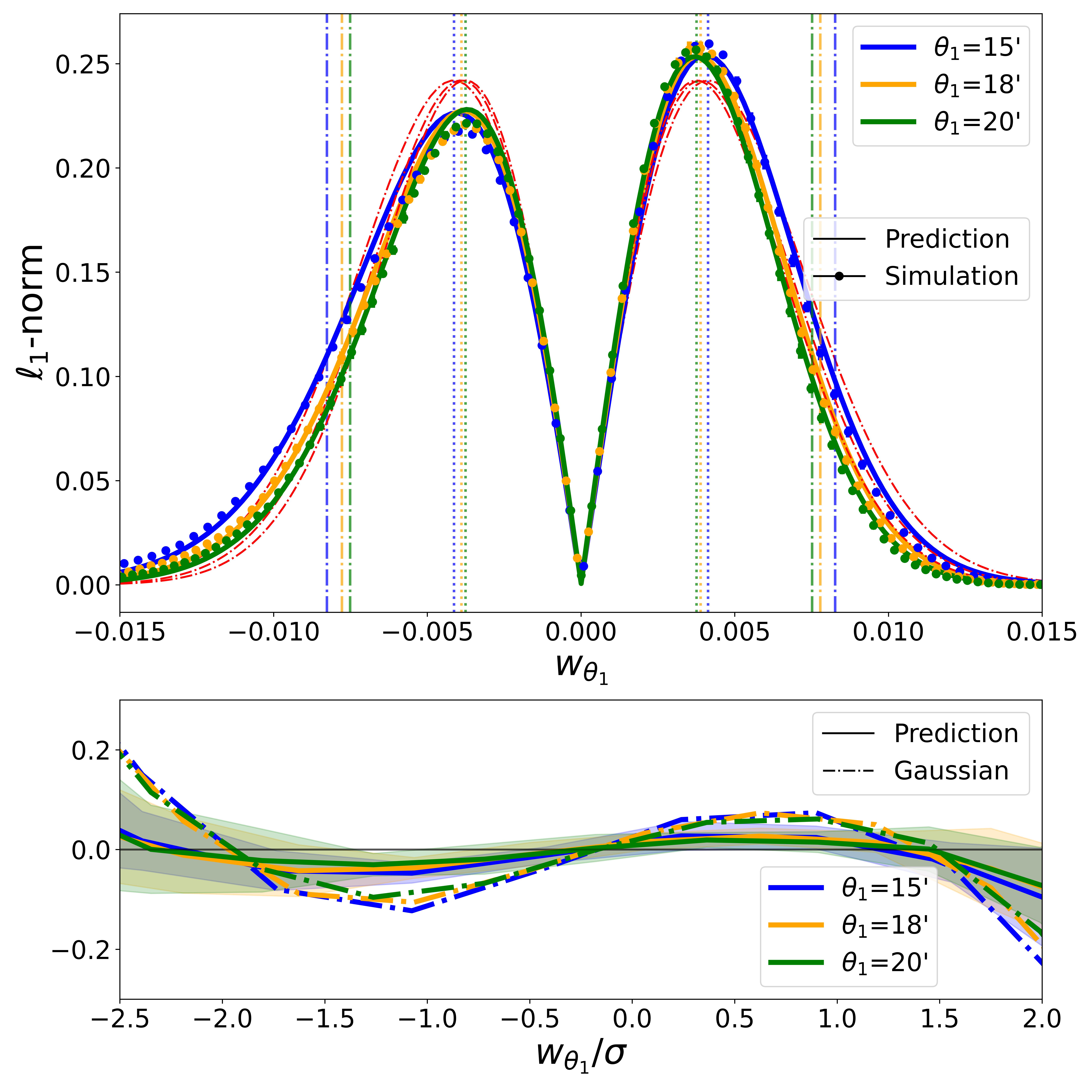}}
      \caption{Top panel: Predicted (solid) $\ell_1$-norm as compared to measurements in simulation (dot markers)for different inner radii $\theta_1 = \SI{15}{\arcminute}, \SI{18}{\arcminute}, \SI{20}{\arcminute}$ and single source redshift $z_s = 1.43$, displayed with blue, orange, and green lines respectively. Red dash-dotted lines show the Gaussian prediction for reference. \vilasini{The vertical dotted and dot-dash lines correspond to the $1\sigma$ and $2\sigma$ regions around the mean of the $w_{\theta_1}$ for each of the case considered.}
      Bottom panel: residual of the prediction relative to the simulation (dotted lines). For reference, the dash-dotted plots illustrate the residual of the $\ell_1$-norm derived from the Gaussian PDF with the same mean and variance as the simulation PDF. The shaded region indicates the $3\sigma$ region around the error bars for each inner \strike{radii} \vilasini{radius}. The prediction is in good agreement with the measurements up to approximately 2 sigma and remains within percent levels.}
         \label{fig: different r}
   \end{figure}

From Table \ref{tab:moments}, we see that the variance from the prediction and simulation for different scales match perfectly, which we expect given the re-scaling by the power spectrum that we perform (see appendix \ref{appendix:a}). As expected from the formalism, we see that the \vilasini{reduced} skewness $(S_3)$ increases with decreasing smoothing scale. We also notice that the $\ell_1$-norm for the positive bins is not at all equal to the negative bins, suggesting how prominent the non-Gaussian features of the mass map are in this regime. It was shown in \cite{2020MNRAS.492.3420B}, that the theoretical precision achievable is constrained by the \strike{blending} \vilasini{mixing} of scales when integrating the density field along the line of sight since highly non-linear scales are not precisely captured by the spherical collapse formalism. Consequently, weak lensing statistical measures are often approached through more phenomenological methods, such as halo models. These models are capable of incorporating baryonic physics, which becomes crucial at smaller scales, as highlighted by \cite{2021MNRAS.502.1401M}. That is also evident when we look at Fig. \ref{fig: different z} and Fig. \ref{fig: different r}, where we do see that the level of accuracy increases as we move to higher redshifts and higher smoothing scales. To bypass this issue, a rising approach is to use the so-called Bernardeau-Nishimichi-Taruya (BNT) transform \citep{2014MNRAS.445.1526B}, which enables \strike{for} more accurate theoretical predictions by narrowing the range of physical scales contributing to the signal.

\subsection{Cosmology dependence of PDF and wavelet $\ell_1$-norm}
In this section, we explore the cosmological dependence of the predicted \strike{aperture mass PDF} \vilasini{pdf of the wavelet coefficients} and $\ell_1$-norm. \vilasini{We would like to emphasize that since the wavelet decomposition is analogous to using an aperture mass map filtering, we henceforth can also call the PDF of the wavelet coefficients as the aperture mass PDF}  From the equation of the rate function given in Eq. \eqref{eq: rate function}, the cosmology dependence of the aperture mass PDF in LDT enters through the scale-dependence of, the non-linear power-spectrum, the dynamics of the spherical collapse and the lensing kernel. Moreover, the presence of massive neutrinos if any also affects the aperture mass as it enters the lensing kernel as given in Eq. \eqref{eq: lensing kernel}, by contributing to the total matter density budget. All of this inevitably means that the summary statistics would also be cosmology-dependent. In the context of applying LDT to predict the wavelet $\ell_1$-norm and PDFs, based on the equations introduced above, we could assume that it should be able to efficiently capture the cosmology dependence. Indeed, the cosmology dependence of the one-\strike{cell} \vilasini{scale} one-point \vilasini{convergence} PDF has been studied in detail in \citet{2021MNRAS.505.2886B} which showed that LDT captures very accurately the cosmology dependence. In this work, we extend this study for the case of \strike{wavelet mass map PDF} \vilasini{the PDF of the wavelet coefficients} and the wavelet $\ell_1$-norm of aperture mass maps. As for the one-\strike{cell} \vilasini{scale} case, the rate function for the multi-\strike{cell} \vilasini{scale} also depends on the variance and the dynamics of the spherical collapse, which means that the \strike{wavelet mass map PDF} \vilasini{PDF of the wavelet coefficients} would also be expected to depend (similarly) on the cosmology. Additionally, since the wavelet $\ell_1$-norm is directly dependent on the PDF, we could naively expect it to depend on the cosmology too, and encapsulate the dependence.

Since we are interested in obtaining the derivatives with respect to cosmological parameters, we need a simulation suite that is available for different cosmologies. For this, we use the Cosmological Massive Neutrino simulations (MassiveNus) \citep{Liu_2018} suite to see how well the theory can capture the dependence on cosmology. These simulations are released by the Columbia Lensing Group \footnote{\url{http://columbialensing.org/}}. The MassiveNuS simulations encompass 101 different cosmological models at source redshifts $z_s = {0.5, 1.0, 1.5, 2.0, 2.5}$, by varying three parameters: the neutrino mass sum $M_\nu$, the total matter density $\Omega_m$, and the primordial power spectrum amplitude $A_s$. For each redshift, there are 10,000 distinct map realizations generated through random rotation and translation of the initial n-body box then stitched together to reconstruct pseudo-independent light cones that are unlikely to cross the same structures. Each $\kappa$ map contains $512^2$ pixels, covering a total angular area of $12.25$ square degrees, spanning a range of $\ell \in [100, 37,000]$ with a pixel resolution of 0.4 arcminutes. The measurement of the wavelet $\ell_1$-norm is obtained using the same process as was described previously in Sec. \ref{sec: takahashi simulation}. Additionally, the measurement of the PDF is derived from the histogram of the binned and smoothed map. To obtain the derivatives, we use the model with parameters $\Omega_m = 0.30$, $A_s = 2.1 \times 10^{-9}$, and $M_\nu = 0.0$ eV as the fiducial model (model 1b in \citet{Liu_2018}) and obtain the derivatives of the PDF for each of the parameters. The different parameters are shown in Table \ref{tab:parameters}. The outcomes are illustrated in Figures \ref{fig:massivenu pdf derivative} to \ref{fig:massivenu l1 derivative}. For the derivative with respect to $M_\nu$, we employ a model with the same $\Omega_m$ and $A_s$ values but with $M_\nu = 0.0, \text{eV}$ and then compute the derivative. 
However, it is important to note that due to resolution and finite volume effects, the simulation power spectrum of the convergence maps exhibits a deficit in power at high $\ell$, which does influence us much because of the scales we are looking at, and at low $\ell$ which invites caution when choosing our largest scales. We typically try to maintain a factor of 20 between the physical scale of the box and the largest physical scale probed by our filters.

It is evident from Figures \ref{fig:massivenu pdf derivative} and \ref{fig:massivenu l1 derivative}, that the prediction effectively captures the changes in cosmology, demonstrating good agreement with the results. \vilasini{The errorbars are obtained by taking the error of mean of 3000 simulations, which are also used to obtain the mean of the measurements.}

\begin{table}
  \caption[]{The cosmological parameters used to obtain the derivatives of the PDF. }
     \label{tab:parameters}
 $$
     \begin{array}{|p{0.3\linewidth}|l|c|r|}

        \hline
        \text{Parameters} & \text{lower bound} & \text{fiducial} &\text{upper bound}\\
        \hline
        $\Omega_m$ & 0.28  &  0.3 & 0.31\\
        $A_s \times 10^{-9}$ &  2.0 & 2.1 & 2.17\\
        $M_\nu$ & 0.09& 0.0 &0.11\\
        \hline
     \end{array}
 $$
\end{table}

\begin{figure}
   \resizebox{\hsize}{!}
            {\includegraphics{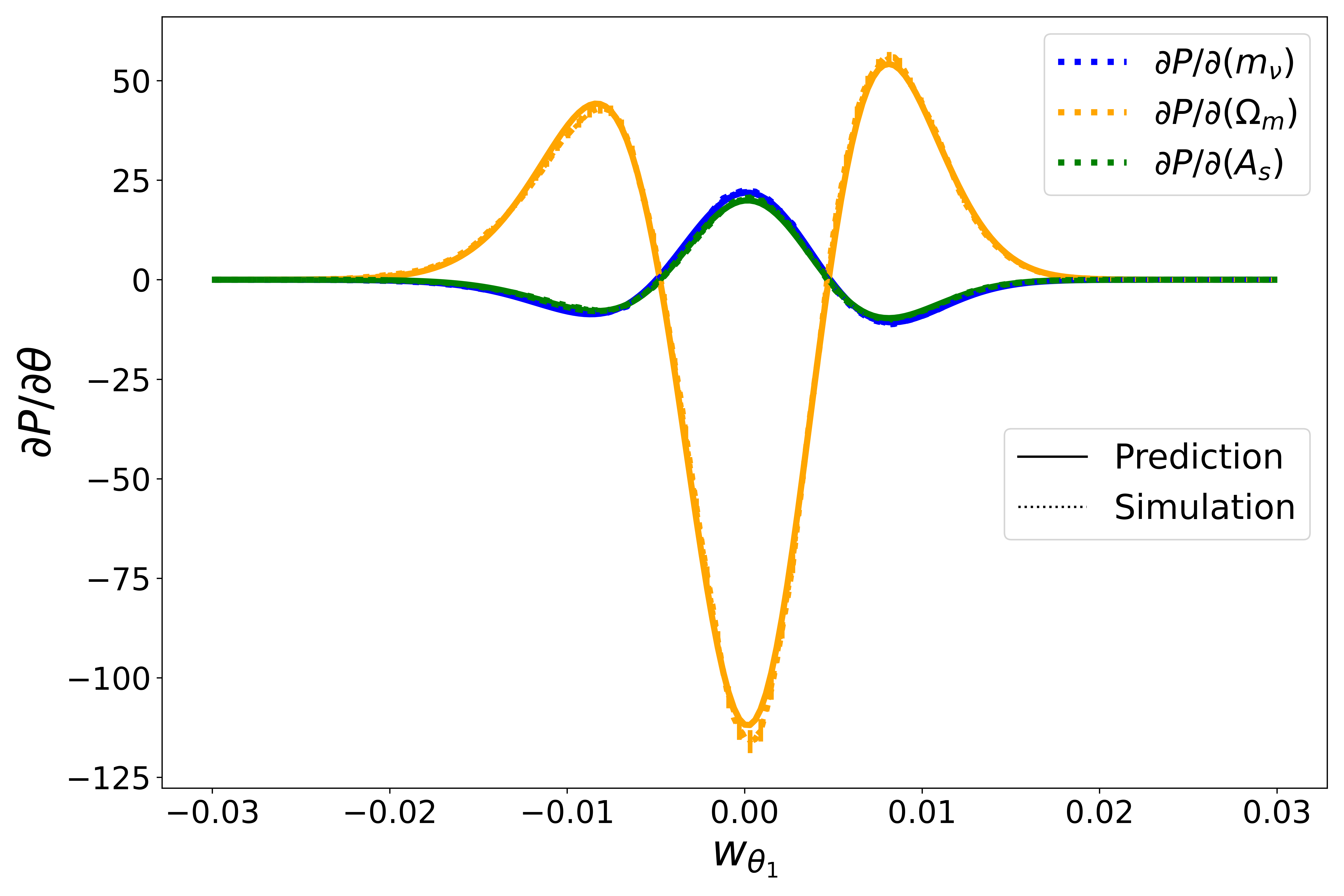}}
    \caption{Derivative of the PDF with respect to $M_\nu$ (blue), $\Omega_m$ (orange) and $A_s$ (green). The solid lines show the derivatives obtained from prediction and the dotted lines show the derivatives from simulation. It is obtained at source redshift $z_s = 2$ and inner radius $\theta_1 =  \SI{22.5}{\arcminute}$. The results for simulation are obtained from the MassiveNus simulation suite, by averaging the results over 10,000 simulations.}
    \label{fig:massivenu pdf derivative}
\end{figure}

\begin{figure}
   \resizebox{\hsize}{!}
            {\includegraphics{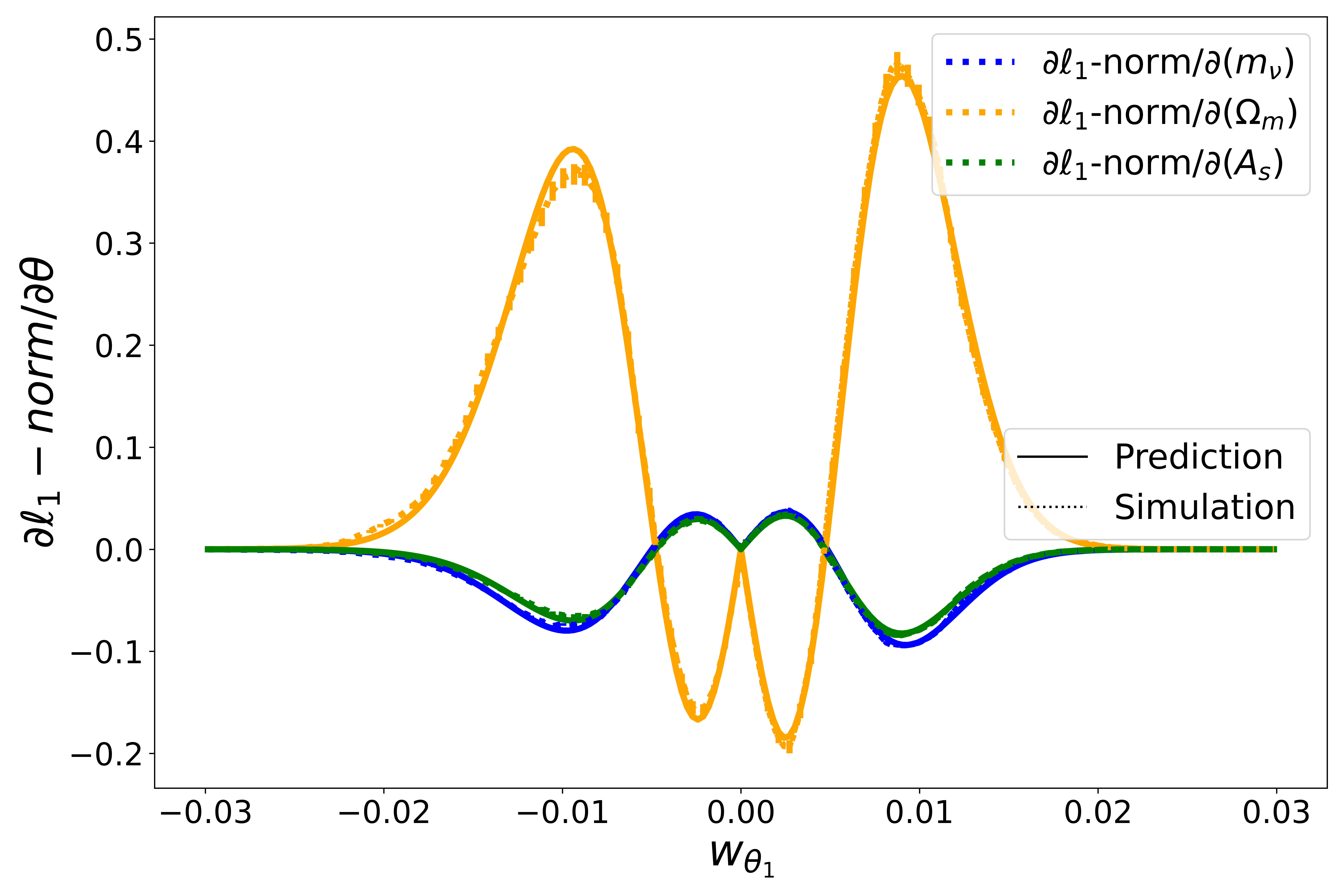}}
    \caption{Derivative of the wavelet $\ell_1$-norm with respect to $M_\nu$ (blue), $\Omega_m$ (orange) and $A_s$ (green). The solid lines show the derivatives obtained from prediction, and the dotted lines show the derivatives from simulation. The predicted wavelet $\ell_1$-norm is derived from the predicted PDF as was explained in previous sections. It is obtained at source redshift $z_s = 2$ and inner radius $\theta_1 =  \SI{22.5}{\arcminute}$. The results for simulation are obtained from the MassiveNus simulation suite, by averaging the results over 10,000 simulations.}
    \label{fig:massivenu l1 derivative}
\end{figure}

As demonstrated in \cite{2021MNRAS.505.2886B} for the case of a 1-cell PDF, our results here demonstrate that the theory captures well the effects of adding massive neutrinos to cosmology, extending the results also for the \strike{aperture mass PDF} \vilasini{PDF of the wavelet coefficients}.

\subsection{Reproducible research}
In the spirit of open research, the code related to reproducing the plots in this paper is available at \href{https://github.com/vilasinits/LDT_2cell_l1_norm}{github}. A part of this Python code that computes the aperture mass/wavelet coefficient PDF was inspired by the public Mathematica code \href{https://github.com/AlexandreBarthelemy/L2DT-Lensing-with-Large-Deviation-Theory}{L2DT} used in \cite{2021MNRAS.503.5204B}.

\section{Discussion and Conclusions}
\label{sec: discussion  and conclusion}
In this work, we have used the \strike{Large Deviation theory} \vilasini{LDT} to build upon the previous work by \citet{2021MNRAS.503.5204B} and introduced a formalism to predict the $\ell_1$-norm of the wavelet coefficients of the \vilasini{lensing}\strike{lesnsing} field and their cosmological dependence for any given redshift and scales as long as they are in the mildly non-linear regime. The approach of \citet{2021MNRAS.503.5204B} incorporates the geometric and time-evolution aspects within the light cone by considering that the correlations of the underlying matter density field along the line of sight are negligible compared to transverse directions. This Limber approximation enables us to treat redshift slices as statistically independent. It was found that the most likely non-linear dynamics of the matter density field filtered in concentric disks can be well approximated for small variances by the cylindrical collapse model, thus allowing for \strike{Large Deviation theory} \vilasini{LDT} to be applied in this context. From there, because the wavelet $\ell_1$-norm can be directly related to the multi-scale one-point statistics of the convergence, we have shown in this paper that a theory-based prediction for the wavelet $\ell_1$-norm is tractable.

The robust validation of our predictions against the simulations, in Figs. \ref{fig: different r} and \ref{fig: different z}, highlighted the reliability and applicability of our theoretical framework. Specifically, our theory aligns with the simulations within percent level accuracy across the examined range of source redshifts $z_s  {\approx 1 - 2}$ and angular scales $\theta_1  {\approx 15-20}$. This quantitative assessment affirms the predictive precision of our model. Obviously going towards smaller scales or lower redshifts would lead to a larger departure between simulations and theory.

Figure \ref{fig:massivenu l1 derivative}, further emphasizes the cosmological dependencies and the very good performance of the theoretical prediction, when compared with that of the simulation. Since the simulations have validated the theoretical predictions, it suggests that our model could be applied to investigate a broad range of cosmological parameters, while bypassing the need for high computing resources and providing a theoretical tool for cosmology inference that is robust and fast, without the loss of quality. This paves the way for a more comprehensive and faster forecast analysis based on theory without having to rely on expensive numerical simulations.
In addition, we noted in Figures \ref{fig:massivenu pdf derivative} and \ref{fig:massivenu l1 derivative} that there is numerical noise in the simulation results (fitted lines). This can lead to biases in parameter estimations when using the simulation data. The use of a theory-based approach has the added benefit of reducing the artificial biases in parameter inference, which was also noted in \citet{2021MNRAS.505.2886B}. 

Another point to be noted is, that although lower source redshifts are expected to generate more non-Gaussian information, this also pushes the model deeper into the nonlinear regime. This effect can be mitigated by adjusting the smoothing scale accordingly. We would also like to emphasize the usefulness of the wavelet coefficients of the convergence maps, instead of the convergence maps or the shear maps directly. Convergence maps already reduce computation expense, because of the presence of a more compressed lensing signal when compared to the shear map. However, the use of the wavelet scales and/or coefficients provides us with a multi-scale analysis method, which is well suited to constrain the cosmology. 

Moreover, wavelet coefficients are invariant under the mass sheet degeneracy, which renders the link to the measured shear data more straightforward.

More generally, one of the issues with the theoretical modelling of weak lensing we currently face is the mixing of (non-linear) scales that is inherent to such quantities projected along the past light cone and makes standard perturbative approaches inaccurate even on relatively large angular scales. \strike{To bypass this issue, a rising approach is to use the so-called Bernardeau-Nishimichi-Taruya (BNT) transform } \citep{2014MNRAS.445.1526B}\strike{, which enables more accurate theoretical predictions by narrowing the range of physical scales contributing to the signal.} \vilasini{To overcome this issue, an emerging method is to utilize the BNT transform which allows for more precise theoretical predictions by reducing the range of physical scales that contribute to the signal.} This could be applied in our context and enables even more accurate predictions for analysing tomographic surveys. Investigating the performance of this approach is left for future works, together with the impact of systematics that would require to devise specific simulated data, as well as additional theoretical development.

We conclude that the novel approach for obtaining the theoretical prediction of the wavelet $\ell_1$-norm summary statistic proposed here presents several advantages in the realm of cosmological parameter inference: i) it provides a fast ($\sim 1$ minute) calculation; ii) it does not rely on a heavy numerical simulation; iii) it captures well the cosmology dependence; and iv) wavelet $\ell_1$-norm could lead to competitive or even tighter constraints as demonstrated by previous works. We hence have at hand a promising method that could enable a comprehensive multi-scale analysis of cosmic shear datasets in the mildly non-linear regime.

\begin{acknowledgements}
       This work was funded by the TITAN ERA Chair project (contract no. 101086741) within the Horizon Europe Framework Program of the European Commission, and the  Agence Nationale de la Recherche (ANR-22-CE31-0014-01 TOSCA and ANR-18-CE31-0009 SPHERES). AB's work is supported by the ORIGINS excellence cluster.

        We thank Jia Liu and the Columbia Lensing group http://columbialensing.org) for making the MassiveNus \citep{Liu_2018} simulations available. 
         The creation of these simulations is supported through grants NSF AST-1210877, NSF AST-140041, and NASA ATP-80NSSC18K1093.

       We thank New Mexico State University (USA) and Instituto de Astrofisica de Andalucia CSIC (Spain) for hosting the Skies \& Universes site for cosmological simulation products.

       We also thank C. Uhlemann, O. Friedrich, Lina Castiblanco, and Martin Kilbinger for insightful discussions.
\end{acknowledgements}

\bibliographystyle{aa} 
\bibliography{references}

\begin{appendix}

\section{Large Deviation Theory}
\label{appendix:a}
\subsection{LDT for the matter field}
The computations detailed in this section draw heavily from the formulations provided in \cite{2021MNRAS.505.2886B} and \cite{2021MNRAS.503.5204B}. Here, we will succinctly restate the key equations and direct the reader to the comprehensive treatments in \cite{2021MNRAS.505.2886B}, \cite{2021MNRAS.503.5204B}, and \cite{Reimberg_2018}.

The \strike{Large Deviations theory} \vilasini{LDT}, as explored in earlier works \cite{doi:10.1137/1.9781611970241}, examines the rate at which the probabilities of specific events diminish as a key parameter of the problem undergoes variation. Widely applied in various mathematical and theoretical physics domains, this theory is particularly prominent in statistical physics, covering both equilibrium and non-equilibrium systems. For a detailed overview, readers are referred to \cite{Touchette_2009}.

The utilization of the Large Deviation Principle (LDP) in the domain of Large Scale Structure cosmology has been systematically developed in recent years and will be applied in the specific context of cosmic shear observations in this work. \cite{2016PhRvD..94f3520B} clarified the application of the theory to the cosmological density field, establishing its link to earlier studies focused on cumulant calculations and modelling the matter \strike{Probability Distribution Function} PDF through perturbation theory and spherical collapse dynamics (\cite{2002A&A...382..412V}; \cite{PhysRevD.90.103519}). \cite{2020MNRAS.492.3420B} provided an LDT-based prediction for the top-hat-filtered weak-lensing convergence PDF on mildly nonlinear scales, building upon earlier work by \cite{2000A&A...364....1B}.

The LDP applied to the matter density field relies on three key aspects:

\begin{itemize}
    \item Defining a rate function for variables in the initial field configuration. In this context, we opt for Gaussian initial conditions and establish their covariance matrix.
    
    \item Describing the relationship between the initial field configuration (representing the mass profile) and the resulting mass profile, based on 2D cylindrical collapse or a suitable approximation.

    \item Using these foundations to express observable quantities, like a map created with a specific filter, as functional expressions that depend on both the final and initial mass profile.
\end{itemize}

A \strike{large deviation principle} \vilasini{LDP} for matter densities at multiple scales (indexed by $i$) $\{\rho^\epsilon_i\}$, $1 \leq i \leq N$ with joint PDF $\mathcal{P}_\epsilon(\{\rho^\epsilon_i\})$ is satisfied if the following limit exists
\begin{equation}
    \psi_{\{\rho^\epsilon_i \}}(\{ \rho^\epsilon_i \}) = - \lim_{\epsilon\to 0} \epsilon \log[\mathcal{P}_\epsilon( \{\rho_i ^\epsilon \} )].
\end{equation}
If such limit exists, $ \psi_{\{\rho^\epsilon_i \}}$ is called the rate function.

Here, $\epsilon$ is a driving parameter, indicating the set of random variables linked to a specific evolutionary process. For the matter density field at a single scale, this parameter reflects its variance, marking time from initial stages to later times. In joint statistics involving concentric disks of matter, the common driving parameter can be selected as the variance at any radius or scale. This consistency arises because all variances behave the same as they approach zero. In this limit, they directly scale with the square of the growth rate of structure in the linear regime.

If LDP holds for the random variable ${\rho_i }$, then Varadhan's theorem allows us to connect its \vilasini{Scaled} Cumulant Generating Function (SCGF) $\varphi_{{ \rho_i}}$ to the rate function  $\psi_{{ \rho_i}}$ through a Legendre-Fenchel transform
\begin{equation}
    \varphi_{\{ \rho_i\}}(\{ \lambda_i\}) = \sup_{\{ \rho_i\}} \Bigl[ \sum_{i}\lambda_i\rho_i -  \psi_{\{ \rho_i\}} (\{ \rho_i\})\Bigr].
    \label{eq: legendre fenchel}
\end{equation}
This Legendre-Fenchel transform reduces to Legendre transform if the rate function is convex
\begin{equation}
    \varphi_{\{ \rho_i\}}(\{ \lambda_i\}) =  \sum_{i}\lambda_i\rho_i -  \psi_{\{ \rho_i\}} (\{ \rho_i\}),
    \label{eq: Legendre transform}
\end{equation}
where $\lambda_i$ and $ \rho_i$ are one by one related via the stationary condition
\begin{equation}
    \lambda_k = \frac{\partial\psi_{\{ \rho_i\}} (\{ \rho_i\}) }{\partial \rho_k}, \forall k \in \{1,\dots,N \} .
    \label{eq: lambda}
\end{equation}

Another consequence of the large-deviation principle is the so-called contraction principle. This principle suggests that when dealing with a set of random variables ${ \tau_i }$ following a \strike{large deviation principle} \vilasini{LDP} and connected to ${\rho_i}$ through the continuous mapping $f$, the rate function of ${\rho_i}$ can be determined as follows
\begin{equation}
    \psi_{\{ \rho_i\}} (\{ \rho_i\}) = \inf_{\{ \tau_i\}:f(\{ \tau_i\}) = \{\rho_i\}} \psi_{\{ \tau_i\}} (\{ \tau_i\}).
\end{equation}
In more tangible terms, this statement suggests that an uncommon change in the behavior of ${\rho_i}$ is predominantly influenced by the most probable variation among all unlikely changes in ${ \tau_i}$.

In the context of cosmology, if we take $\rho_k$ to denote the late-time densities, then $\bar{\tau_k}$, the most likely initial field configuration, is obtained through the most probable mapping between the linear and late-time fields. In 2D, this most likely dynamics is given by cylindrical collapse, which is known to be well-fitted by
\begin{equation}
\zeta(\bar{\tau_k}) = \rho_k = \Bigl( 1 - \frac{\bar{\tau_k}}{\nu}\Bigr)^{-\nu}
\label{eq: spherical collapse}.
\end{equation}
The choice of $\nu$ as 1.4 is made to match the calculated value of the third-order skewness of the matter density contrast in cylinders from perturbation theory, as presented in \cite{2018MNRAS.477.2772U}. The standard result for spherical collapse dynamics in 1 to 3D can be found in \cite{Mukhanov}.

Now that we know the most likely connection between initial conditions and the late-time field configuration, we can compute the rate function for the late-time density field. It is expressed as follows
\begin{equation}
\psi ({ \rho_i}) = \frac{\sigma^2_{R1}}{2} \sum_{k,j}\Xi_{kj}({\tau_i }) \bar{\tau_k} \bar{\tau_j}.
\label{eq: rate function}
\end{equation}

Here, $\sigma^2_{R_1}$ is the driving parameter and is determined by the variance within the largest disk $R_1$. $\Xi_{kj}$ is the inverse covariance matrix between the linear density field inside the initial disks of radii $R_k\rho_k^{1/2}$.

\subsection{Applying LDT to aperture mass PDF}
Following \cite{2021MNRAS.503.5204B}, an accurate modelling of the top-hat smoothed weak lensing convergence can be obtained (Limber) approximating that the convergence field is an assembly of statistically independent infinitely long cylinders of the underlying matter density contrast. Those cylinders are centred on slices along the line of sight, and they reduce to those 2D slices as illustrated in Fig. \ref{fig: lensing_planes}. This means that the cumulants can be written as
\begin{figure}
   \resizebox{\hsize}{!}
            {\includegraphics{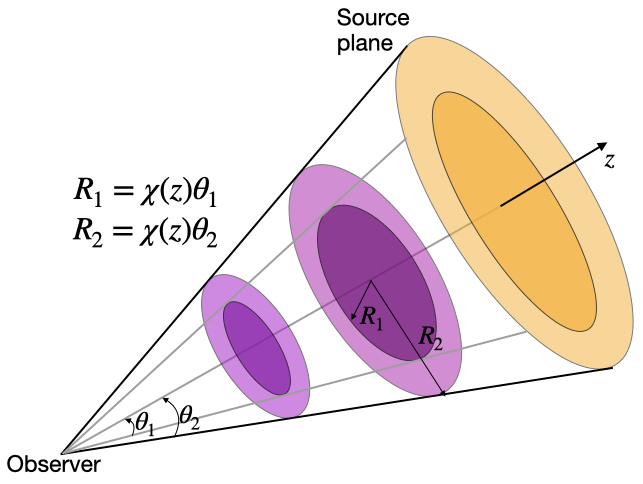}}
      \caption{Schematic view of the procedure to obtain the aperture mass map following \citep{2021MNRAS.503.5204B}. The projected quantities can be inferred as a superposition of the underlying 3D density field along the line of sight.}
         \label{fig: lensing_planes}
   \end{figure}
\begin{equation}
    \bigl< M_{ap}^p\bigr>_c = \int_{0}^{\chi_s} d\chi \vilasini{\omega(\chi, \chi_s)}\bigl<( \delta_{<\mathcal{D}(\chi)\theta_2} - \delta_{<\mathcal{D}(\chi)\theta_1})^p \bigr>_c,
    \label{eq: Map cumulant}
\end{equation}
where $\delta_{<\mathcal{D}(\chi)\theta_2} - \delta_{<\mathcal{D}(\chi)\theta_1}$ is a random variable that defines the slope between two concentric disks of radii $\mathcal{D}(\chi)\theta_2$ and $\mathcal{D}(\chi)\theta_1$ at comoving radial distance $\chi$. This simplifies the problem greatly, now having to calculate just the one-point statistics related to the density slope within each two-dimensional slice along the line of sight.

As is explained in Appendix A of \cite{2021MNRAS.503.5204B}, the choice of driving parameter is not predicted by theory and is left as a free parameter. However, when computing the joint statistics of the density fields at different scales, this choice prevents us from imposing correct quadratic contributions in the CGF. This leads us to use the full non-linear prescription coming from the Halofit to model the non-linear covariance. In this paper, we use the Halofit-Takahashi version \cite{Takahashi_2012} to compute the covariances.

Now using Eq. \eqref{eq: Legendre transform}, and Eq. \eqref{eq: rate function}, and since we need the density slope between two concentric disks, we get the \vilasini{S}CGF as
\begin{equation}
    \varphi_{\delta_2 - \delta_1}(\lambda) = \varphi_{\delta_1, \delta_2}(-\lambda, \lambda)
    \label{eq: CGF}
\end{equation}

To ensure that this choice does not lead to any discrepancies with the numerical simulations, we also rescale the projected CGF given in Eq. \eqref{eq: Legendre transform} with the measured variance $\sigma_{M_{ap},sim}^2$ instead of the one computed from Halofit  $\sigma_{M_{ap},hfit}^2$ as is given below
\begin{equation}
    \phi_{M_{ap}}(\lambda) = \frac{ \sigma_{M_{ap},hfit}^2}{ \sigma_{M_{ap},sim}^2} \phi_{M_{ap}} \left(\lambda \frac{ \sigma_{M_{ap},sim}^2}{ \sigma_{M_{ap},hfit}^2} \right )
\end{equation}

Once the CGF of individual slices is obtained, we can use Eq. \eqref{eq: Map cumulant} to get the CGF \vilasini{(the details of this derivation is given in \cite{2021MNRAS.503.5204B})} of the lensing aperture mass
as
\begin{equation}
    \phi_{\kappa,\theta_1, \theta_2}(\lambda) = \int_{0}^{\chi_s}d\chi w^p(\chi,\chi_s) \phi_{\delta1,\delta_2,M_{ap}}(w(\chi,\chi_s)\lambda, \mathcal{D}(\chi)\theta1, \mathcal{D}(\chi)\theta_2).
    \label{eq: SCGF}
\end{equation}

After calculating the CGF, the inverse Laplace transform can be used to obtain the convergence PDF
\begin{equation}
    P(\kappa) = \int_{-i\infty}^{+i\infty} \frac{d\lambda}{2\pi i} \exp(-\lambda \kappa + \phi_{\kappa,\theta}(\lambda)).
    \label{eq: inverse laplace}
\end{equation}

As described in more detail in \cite{2021MNRAS.503.5204B}, the inverse Laplace transform we need to perform assumes that the CGF is defined in the complex plane along the path of integration. Unfortunately, the use of numerical results for the matter power spectra prevents us to perform this continuation from the real axis. As a result, we use an informed fit of the numerical CGF along the real axis with a finite number of coefficients that we then extend to the complex plane. This allows to perform the previous integral.

Now that we have a PDF of the aperture mass map $P(M_{ap})$/ alternatively the wavelet coefficients, we could use that to obtain the wavelet $\ell_1$-norm using the Eq. \eqref{eq: derived l1 norm} as demonstrated in the Sect. \ref{sec: deriving l1 norm from pdf}.

\end{appendix}

\end{document}